\documentclass[a4paper]{jpconf}

\usepackage{graphicx}

\usepackage{amssymb}

\begin{document}
\title{Acceleration and propagation of ultrahigh energy cosmic rays}

\author{Martin Lemoine}
\address{Institut d'Astrophysique de Paris,\\ CNRS, UPMC, \\ 98 bis
  boulevard Arago, F-75014 Paris, France}
\ead{lemoine@iap.fr}

\begin{abstract}
  The origin of the highest energy cosmic rays represents one of the
  most conspicuous enigmas of modern astrophysics, in spite of
  gigantic experimental efforts in the past fifty years, and of active
  theoretical research. The past decade has known exciting
  experimental results, most particularly the detection of a cut-off
  at the expected position for the long sought Greisen-Zatsepin-Kuzmin
  suppression as well as evidence for large scale anisotropies. This
  paper summarizes and discusses recent achievements in this field.
\end{abstract}

\section{Introduction}
Ultrahigh energy cosmic rays are particles with energies $\gtrsim
10^{18}\,$eV that form the end of the cosmic ray spectrum. What they
are and where they come from remain long-standing questions, the first
detection of a $10^{20}\,$eV air shower at Volcano Ranch~\cite{L63}
dating back to some fifty years ago. Larger and larger detectors have
been built ever since, culminating today at the scale of thousands of
km$^2$; and yet, this field remains data driven or actually, data
starved. The extraordinarily low flux at these energies certainly
implies low statistics, of the order of 1 particle per km$^2$ and per
century at $\sim 10^{20}\,$eV. As a direct consequence, the world
catalogue contains only a dozen or so of events with energy $E\gtrsim
10^{20}\,$eV.

It is generally admitted that ultrahigh energy cosmic rays originate
in extragalactic sources, mainly because light nuclei above
$10^{18}-10^{19}\,$eV cannot be confined in the Galactic magnetic
field and because the arrival directions of ultrahigh energy cosmic
rays are essentially isotropic. On top of that, only a few types of
objects appear to be capable of accelerating particles to such extreme
energies, and for such sources, the extragalactic flux is expected to
dominate.

In such a picture, one is tempted to attribute the spectral break of
the cosmic ray spectrum at $\sim 10^{19}\,$eV (the so-called ankle) to
the energy at which the ultrahigh energy component emerges on top of a
lower energy component. However, one cannot exclude that cosmic rays
from the ``second knee'' at $\sim10^{17}-10^{18}\,$eV upwards form a
single extragalactic component; in such a scenario, assuming that
cosmic rays around the ankle are essentially protons, the ankle
feature itself may well result from pair production of ultrahigh
energy protons on the cosmic microwave background~\cite{BG88,Bea06}.
Where ``second knee'' cosmic rays originate from and at which energy
the extragalactic flux steps over the Galactic component also remain
open questions; nevertheless the following discussion concentrates on
extragalactic cosmic rays of energy $\gtrsim
10^{18}\,$eV. Section~\ref{sec:qu} discusses the main questions and
problems, briefly summarizing at the same time the current
experimental results, some of which are covered in greater detail by
other papers in this volume. Section~\ref{sec:acc} focusses on the
difficulties of accelerating particles to such energies,
Section~\ref{sec:prop} discusses some aspects of propagation and
comments on the observed anisotropies. Section~\ref{sec:concl} draws
conclusions and briefly discusses some perspectives in this field.

\section{Many questions... a few hints}\label{sec:qu}
The crucial questions and salient recent experimental results in this
field can be categorized as follows:

\subsection{Where does the cosmic ray spectrum stop?}
One expects the spectrum to cut-off short of $10^{20}\,$eV, as the
Universe becomes increasingly opaque to nuclei at these
energies~\cite{G66,ZK66} (GZK). For protons, this GZK suppression
results from photo-pion interactions $p+\gamma\rightarrow N+\pi$ (with
$N=p,n$ a nucleon) on cosmic microwave background photons, which
induce energy losses $\sim m_\pi/m_p\sim 15\,$\% every interaction
length $\sim 10\,$Mpc above the threshold $\sim 6\times 10^{19}\,$eV,
thereby leading to an energy loss length $\sim50-100\,$Mpc. More
quantitatively, 90\% of protons with recorded energy $\geq10^{20}\,$eV
have travelled less than $130\,$Mpc, while 90\% of protons with energy
$\geq 6\times 10^{19}\,$eV come from distances smaller than
200~Mpc~\cite{HMR06}.

Similar conclusions apply for nuclei with charge $Z>1$, although the
relevant processes are photodisintegration interactions stripping one
or a few nucleons off the nucleus, through the interaction with
infrared or microwave background photons. The energy loss length of
iron nuclei is roughly comparable to that of protons, i.e. 90\% of
iron nuclei with energy above $10^{20}\,$eV originate from less than
$70\,$Mpc, while 90\% of iron nuclei with energy above $6\times
10^{19}\,$eV originate from less than $180\,$Mpc~\cite{HMR06}. The
energy loss length of smaller atomic number species is smaller, all
the more so as the charge becomes smaller; this effect results from
the scaling of the threshold energy for photodisintegration $\propto
A$.  Helium nuclei with $E\gtrsim 6\times 10^{19}\,$eV are thus
photodisintegrated into nucleons within a few Mpc from the source, see
e.g. ~\cite{Bea02,Alea08}.

\begin{figure}
\includegraphics[width=23pc]{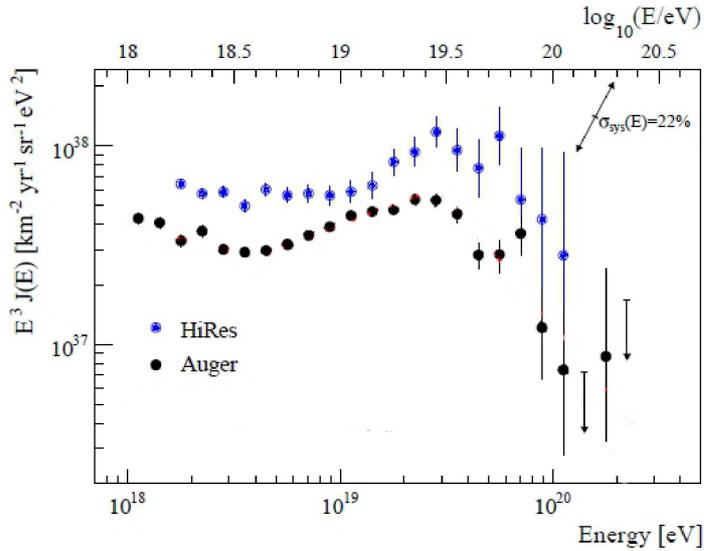}
\begin{minipage}[b]{14pc}
  \caption{Differential spectrum of ultrahigh energy cosmic rays
    (multiplied by $E^3$) as measured by the HiRes experiment and the
    Pierre Auger Observatory, revealing the cut-off at the expected
    location for the GZK effect. Adapted from
    ~\cite{D10}.\label{fig:hires-spec} }
\end{minipage}
\end{figure}

The first detection of a flux suppression above $6\times 10^{19}\,$eV
has been reported by the HiRes experiment~\cite{Aea08}. This
suppression has been confirmed by the Pierre Auger
Observatory~\cite{Aea08PAO} and more recently by the Telescope Array
experiment~\cite{TA12}. This feature appears clearly in the spectra
measured by HiRes and by the Pierre Auger Observatory, shown in
Fig.~\ref{fig:hires-spec}. Note that the apparent discrepancy between
these two spectra lies at the level of the systematic uncertainties on
energy calibration, of order $20\,$\%.  This discovery is in
remarkable agreement with the assumption of an extragalactic ultrahigh
energy cosmic ray proton spectrum. However, one must keep in mind
that: (1) mixed composition spectra, with a composition for instance
similar to that of Galactic cosmic rays may provide a roughly similar
cut-off; (2) a maximal acceleration energy in the source at
$\sim6\times 10^{19}\,$eV may lead to a similar propagated spectrum,
even for sources located in our vicinity.  Nevertheless, if one
attributes this cut-off to the GZK effect, one infers that the sources
of particles above $6\times 10^{19}\,$eV are extragalactic and lie
within 100Mpc. This is our cosmological backyard, so to speak.

\subsection{What are ultrahigh energy cosmic rays?}
This question is particularly crucial because, depending on the charge
of the particle, one expects widely different phenomenologies. In
particular, a particle of charge $Z$ can generally be accelerated to
$Z$ times the energy of a proton but, at the same time, it is
deflected by an angle $Z$ times that of a proton in intergalactic and
Galactic magnetic fields.

Unfortunately, the experimental situation is rather confuse in this
respect because various experiments suggest different compositions. So
far, the most reliable measurements of chemical composition have come
through the determination of the maximum of shower development in the
atmosphere, written $X_{\rm max}$ and expressed in grammage g/cm$^2$,
which is measured by the fluorescence telescopes that follow the
longitudinal development of the air shower.  The average $\langle
X_{\rm max}\rangle$ measured by the Fly's Eye experiment has provided
indication early on that the chemical composition turns from heavy
below the ankle to light above~\cite{Bea93}. The HiRes measurements,
which recover the previous Fly's Eye results where applicable,
indicate a light composition up to $\sim10^{19.7}\,$eV~\cite{Aea10}.
Recently the Pierre Auger experiment has reported measurements of
$\langle X_{\rm max}\rangle$ that point toward a light composition
close to the ankle, which becomes increasingly heavier as the energy
is increased~\cite{Aea10PAO}. This measurement is further supported by
the trend of the fluctuations of $X_{\rm max}$ from shower to shower
with energy. There is still some residual discrepancy between these
various indicators in the Pierre Auger data, as the fluctuations tend
to suggest a heavier composition than the mean $X_{\rm max}$, but the
discrepancy remains marginal at this stage.  Finally, the most recent
data of the Telescope Array, located in the Northern hemisphere as
HiRes, also suggest a proton dominated composition up to
$10^{19.7}\,$eV~\cite{S10}.

This situation is all the more frustrating when one realizes that the
differences between the $X_{\rm max}$ values of all three experiments
lie within the experimental uncertainties. Thus, one cannot draw any
clear information at this stage. 

As emphasized below, the whole phenomenology and the interpretation of
current data, as well as the prospects for the future, differ widely
depending on the composition. At the highest energies, above the GZK
cut-off, one would expect the composition to be dominated by iron
group nuclei or by protons, since intermediate mass nuclei should have
photodisintegrated on smaller distance scales, as discussed
above. This statement implicitly assumes that all sources are equal:
then, the flux received from sources out to distance $d$ scales as
$d$, so that most of the flux is received from sources located close
to the horizon for particles of a given energy; as intermediate mass
nuclei disappear on a length scale that is significantly smaller than
the horizon of heavier nuclei or of protons, these two latter should
dominate the composition at detection. One thus generally consider
these two extremes, proton vs. iron group nuclei, as representative
cases of differing chemical compositions. At energies below GZK, one
may of course envisage that the chemical composition be more complex.

To the above two questions, one should add the following two: ``what
kind of source accelerates particles to such extreme energies?'', and
``why do we not see the sources in the arrival directions of the
highest energy cosmic rays?''. The two sections that follow are
dedicated to these questions, which form the basis of most of the
theoretical activity in this field.

\section{Acceleration to ultra-high energies}\label{sec:acc}
Turning to theory for further clues, one should first address the
problem of accelerating particles up to the extreme energies. The
well-known Hillas plot~\cite{H84} draws a list of possible sources of
the highest energy cosmic rays by making the simple yet efficient
statement that during acceleration the particles must be confined in
the source on a Larmor timescale, i.e $r_{\rm L}\leq R$, with $R$ the
size of the source and $r_{\rm L}=\epsilon/(Ze B)$ the Larmor radius
of the particle of energy $\epsilon$ in the source magnetic field
$B$. This criterion is well designed for non-relativistic sources but
relativistic effects deserve special attention, see below. The Hillas
plot provides a necessary but not sufficient condition for
acceleration to a given energy. To make further progress, one needs to
compare the acceleration timescale $t_{\rm acc}$ to all relevant
timescales, e.g. ~\cite{NMA95,Hea99}. Then, one concludes that the
leading contenders for accelerating protons to the highest energies
are the most powerful radiogalaxies~\cite{T90,RB93}, far away from the
blazar zone where radiative losses prevent acceleration to ultrahigh
energies, gamma-ray bursts~\cite{MU95,V95,W95} and young fast spinning
magnetars~\cite{BEO00,A03,FKO12}.

\subsection{General considerations}\label{sec:genacc}
One often assumes, out of simplicity, that the acceleration timescale
$t_{\rm acc}={\cal A}r_{\rm L}/c$ with ${\cal A}\sim 1$. This, however
is most often too optimistic. Such a statement is also frame dependent
and as such it must be treated with caution.  Particle acceleration
requires electric fields $E$ and in such electric fields, the equation
of motion $\mathbf{p}\cdot{\rm d}\mathbf{p}/{\rm d}t= q_e\,
\mathbf{p}\cdot\mathbf{E}$ indicates that acceleration occurs (of
course) as long as the particle is moving along (or against, if
$q_e<0$) the electric field; then, the acceleration timescale $t_{\rm
  acc}\,\sim\, pc/(eE)$. However $E\,\leq\,B$ in great generality,
since Ohm's law applied to the near perfectly conducting astrophysical
plasmas implies $\mathbf{E} + \mathbf{v}\times\mathbf{B}/c\,\simeq\,0$
in a plasma moving at velocity $\mathbf{v}$; thus, at the very least
${\cal A}\sim c/\vert\mathbf{v}\vert$ (see also ~\cite{LO07} for a
detailed discussion). More importantly, drifting in a homogeneous
coherent magnetic field does not lead to energy gain: in order to
experience the voltage, particles must travel across the field
lines. How such cross-field transport is realized often differentiates
one acceleration mechanism from another and controls the acceleration
timescale, more than often pushing ${\cal A}$ to values larger or much
larger than the above.

For example, non-relativistic diffusive shock acceleration leads to
${\cal A}\sim \beta_{\rm sh}^{-2}t_{\rm scatt}/t_{\rm L}\,\gg\,1$
because the energy gain per cycle around the shock front $\Delta
E/E\sim \beta_{\rm sh}$ but the cycle timescale $\sim t_{\rm
  scatt}/\beta_{\rm sh}$ ($\beta_{\rm sh}= v_{\rm sh}/c\ll1$ the shock
velocity in units of $c$)~\cite{BE87}.  The scattering timescale
$t_{\rm scatt}$ is generically larger, possibly much larger than the
gyration timescale $t_{\rm L}\equiv r_{\rm L}/c$,
e.g.~\cite{P01,CLP02}. In stochastic acceleration, as in the original
Fermi mechanism~\cite{F49}, the particle diffuses in momentum space
with a typical acceleration timescale corresponding to ${\cal A}\sim
\beta_*^{-2}t_{\rm scatt}/t_{\rm L}$, where $\beta_*$ represents the
scattering centers velocity in units of $c$. In a similar process, a
particle may gain energy from the electric component of turbulent
waves, but $\vert E_k\vert/\vert B_k\vert=\omega_k/(kc)\sim \beta_{\rm
  A}$ for Alfv\'en waves of velocity $\beta_{\rm A}c$. In this case
therefore, ${\cal A}\sim \beta_{\rm A}^{-2}t_{\rm scatt}/t_{\rm
  L}\,\gg\,1$~\cite{OSRT09}.

All things being equal, acceleration to extremely high rigidities thus
requires relativistically moving scattering centers. If the scattering
centers move as fast as the particle that is being accelerated, subtle
effects come into play, see ~\cite{P99} for the discussion of
stochastic acceleration and ~\cite{GA99} for relativistic Fermi
acceleration.  It is fair to say that both schemes are not yet fully
understood at the present time.

Relativistic shock acceleration deserves a special focus as it has
received a great deal of attention and because it finds a natural
setting in relativistic outflows. Analytical and numerical studies
indicate that relativistic Fermi acceleration can be operational only
at weakly magnetized shock waves, see ~\cite{LP10,LP11} and references
therein. The maximal energy is limited by the fact that the particle
interacts with self-generated turbulence at the microscopic scale,
which is rather inefficient in terms of scattering high energy
particles, meaning $t_{\rm scatt}\,\gg\,t_{\rm L}$. As discussed
in~\cite{LP11,PPL12}, a reasonable maximum energy estimate is obtained
by balancing the upstream residence time in the background unamplified
magnetic field $B_0$, with the age $r/c$ ($r$ expansion radius) of the
shock wave, leading to $E_{\rm max} \,\sim\,Z \gamma_{\rm sh} e B_0 r$
($\gamma_{\rm sh}$ shock Lorentz factor). This is $\gamma_{\rm sh}$
times larger than the naive Hillas estimate, but it is to be
calculated in the background unamplified field, not in the
self-generated turbulence. It therefore leads to a rather
disappointing $E_{\rm max} \,\sim\,Z \times 3\times 10^{15}\,{\rm
  eV}\,\gamma_{{\rm sh},2} B_{-6} r_{17}$ with $\gamma_{\rm sh,2}\equiv
\gamma_{\rm sh}/100$, $B_{-6}\equiv B/1\,\mu$G, $r_{17}\equiv
r/10^{17}\,$cm.  In short, ultra-relativistic shocks are not efficient
accelerators to ultra-high energies.

Mildly relativistic shocks thus emerge as the most promising Fermi
accelerators, with $t_{\rm acc}\sim t_{\rm scatt}$.  Achieving $t_{\rm
  acc}\sim t_{\rm L}$ requires the efficient generation of turbulence
in the upstream with a nearly scale invariant spectrum of turbulence;
whether this can be realized or not remains to be determined.

There exist of course other possible acceleration mechanisms.  In
magnetized rotator (pulsar) models for instance, cross-field transport
to ultrahigh energies may be achieved through the ponderomotive force
exerted by electromagnetic waves on the particles -- provided these
waves retain their coherence -- or through inertia effects in the
accelerating wind~\cite{A03}. In rotating magnetospheres of neutrons
stars and black holes, the main problem is how to tap the huge voltage
-- e.g. $\Phi\sim 10^{22}V B_{15}R_{6}^3P_{\rm msec}^{-2}$ for a
$P_{\rm msec}$ millisecond period magnetar with surface dipolar field
$B=10^{15}\,B_{15}\,$G and radius $R=10^6R_6\,$cm -- and how to avoid
catastrophic energy losses in the rather harmful environment,
see~\cite{A03,Bea90}. Centrifugal acceleration in black hole
magnetospheres operates thanks to inertia effects, which produce a
drift in the direction of the convective electric field; however, the
energy gain may not be sufficient to reach the ultra-high energy
regime~\cite{RA09,IS09}. Wave particle interactions have also been
suggested as a mechanism to push the particles along $\mathbf{E}$ in
powerful sheared jets~\cite{LO07}. Shear acceleration represents an
interesting variant of the Fermi process that exploits the velocity
gradient of a flow to generate the motional electric
fields~\cite{RBRD07}, and which may lead to ultrahigh energy cosmic
ray production in gamma-ray burst jets~\cite{RD05}. In reconnection
sites, the maximal energy is limited by the extent $l_{\rm rec}$ of
the reconnecting layer, $E_{\rm max}\,\sim\, \beta_{\rm rec} Z e B
l_{\rm rec}$ ($\beta_{\rm rec}$ velocity of the reconnection process
in units of $c$), but under certain assumptions, this might lead to
$10^{20}\,$eV particles in highly magnetized jets~\cite{G10}.

\subsection{Candidate sites}\label{sec:cand}
Gamma-ray bursts, magnetars and powerful radio-galaxies generally rank
as the top three candidates. For magnetars and/or young pulsars, see
~\cite{BEO00,A03,FKO12}. The expected composition is unknown but, if
the particles are to be stripped off the surface of the star, a light
composition appears more likely ~\cite{A03}. Nevertheless, assuming
an ad-hoc mixed composition at the source with a hard spectrum,
~\cite{FKO12} show that the energy losses in the surrounding supernova
enveloppe might lead to a relatively soft spectrum with a light to
heavy trend with increasing energy, in satisfactory agreement with the Pierre
Auger data.

Quite a few acceleration processes have been proposed in the
literature to account for ultrahigh energy cosmic ray acceleration in
gamma-ray bursts, e.g. acceleration in internal shocks~\cite{W95}, at
the external shock~\cite{V95}, through unipolar induction~\cite{MU96},
through stochastic interactions with internal shock
fronts~\cite{GP04}, at the reverse shock~\cite{W01,DR10}, through
stochastic acceleration behind the forward external shock~\cite{DH01},
through shear acceleration~\cite{RD05}, or reconnection~\cite{G10}.
Interestingly, if acceleration takes place in the internal shock
phase, one may expect a strong neutrino signature due to proton
interactions with the radiative background~\cite{WB97,W01}. Such a
signature is now being probed by the Ice Cube experiment~\cite{IC12},
see also ~\cite{HBW12,Hea12}. The possibility of detecting radiative
signatures of proton acceleration to ultra-high energies has also been
discussed~\cite{AIM09}. One generally expects a light composition
although heavy nuclei remain a possibility if they survive
photodisintegration and spallation within the
flow~\cite{L02,Wea08}. In any case, the strongest limitation remains
the energy injected in ultra-high energy cosmic rays: gamma-ray bursts
should indeed produce an isotropic equivalent $\sim10^{53}\,$ergs in
cosmic rays above $10^{19}\,$eV assuming a present day rate $\dot
n_{\rm GRB}\sim1\,$Gpc$^{-3}$yr$^{-1}$ in order to match the observed
cosmic ray flux. This is larger than the typical output energy in
photons by roughly an order of magnitude, see~\cite{Kea09,EP10} for
detailed estimates. Low-luminosity gamma-ray bursts, which are much
more numerous, but also less poweful, represent an interesting
alternative, e.g.~\cite{Wea07,Lea11,LW12}. Such sources can in
principle push nuclei up to $\sim Z\times10^{19}\,$eV, so that one
would expect the composition to be C,N,O dominated close to the GZK
cut-off.

In a way, active galactic nuclei meet opposite difficulties than
gamma-ray bursts: the injection of only a moderate fraction of their
bolometric luminosity would suffice to repoduce the observed cosmic
ray flux above $10^{19}\,$eV~\cite{Bea06,PRZ11}, yet they do not all
offer as appetizing physical conditions with respect to particle
acceleration as gamma-ray bursts. Enormous luminosities are actually
required to accelerate protons to the highest energies observed, which
limits the potential number of active sources in the nearby Universe
(unless the highest energy cosmic rays are heavy nuclei of course).
To see this, consider that acceleration takes place in an outflow,
which may be moving at relativistic velocities or not, with bulk
Lorentz factor $\gamma$, and write the acceleration timescale as
before, $t_{\rm acc}\,=\,{\cal A}t_{\rm L}$ (see also
~\cite{NMA95,LO07,W05,LW09,DR10,MDTM12}). In the comoving frame, the
maximal energy is at least limited by the condition $t_{\rm
  acc}\,<\,t_{\rm dyn}=R/(\gamma\beta c)$, with $R$ the distance to
the origin the outflow, the quantity $t_{\rm dyn}$ defining the
dynamical timescale. This inequality can be rewritten as a lower bound
on the magnetic luminosity of the source, which is defined as
$L_B\,=\,R^2\Theta^2 \gamma^2\beta c B^2/4$ in the laboratory or
source frame:
\begin{equation}
L_B\,\geq\, 0.65\times 10^{45}\, \Theta^2\gamma^2{\cal
  A}^2\beta^3Z^{-2}E_{20}^2\, {\rm erg/s}\ ,\label{eq:LB}
\end{equation}
with $E_{20}$ the observed energy in units of $10^{20}\,$eV, $\Theta$
the opening angle of the outflow and $\beta$ the outflow velocity in
units of $c$. Recall now that ${\cal A}\gtrsim1$ in generic
acceleration processes, and often ${\cal A}\gg1$. For protons,
Eq.~(\ref{eq:LB}) leads to a limit that only a few sources can
achieve. To be more precise, in the compilation of leptonic models
by~\cite{CG08}, only the most powerful flat spectrum radio quasars,
which are though to be the jet-on analogs of FR~II radio-galaxies with
relativistic jets, show a magnetic luminosity in excess of
$10^{45}\,$erg/s. BL Lac objects or TeV blazars, thought to be the
analogs of FR~I radio-galaxies (such as Cen~A) typically exhibit
magnetic luminosities of the order of $10^{44}\,$erg/s or less. Recall
furthermore that the above does not consider the possible radiative
losses in the blazar zone, which further degrade $E_{\rm
  max}$~\cite{NMA95,Hea99,MDTM12}. Shocks in the jets and the hot
spots of the most powerful FRII radio-galaxies may nevertheless offer
the requisite conditions~\cite{T90,RB93}.

Whether one is dealing with protons or nuclei leads to very different
conclusions, all the more so as $Z\gg1$. For $Z\sim 26$, in
particular, the pool of possible candidates blows up. Of course, one
expects a mostly light composition on the basis of the low universal
abundance of heavy nuclei; however, if protons are accelerated up to
some sub-GZK energy $E_p$, nuclei of charge $Z$ may well dominate
beyond, up to $ZE_p$~\cite{Alea08,Aea12}. Could the paucity of proton
FR~II sources in the GZK sphere (radius $\sim 100$Mpc) be compensated
by the acceleration of heavier nuclei in the less powerful and more
numerous FR~I radio-galaxies?  Ptuskin and collaborators have shown
that if all radio-galaxies inject a mix of light to heavy elements
with a rigidity dependent maximal energy following Eq.~(\ref{eq:LB}),
$L_B$ being scaled to the radio luminosity, then indeed one could
explain rather satisfactorily the observed
spectrum~\cite{PRZ11}. However, as discussed in the following, it
becomes very difficult in such scenarios to understand the observed
pattern of anisotropy if one assumes that the highest energy particles
are heavier than hydrogen~\cite{LW09}.

\section{Transport of ultrahigh energy cosmic rays}\label{sec:prop}

The arrival directions are roughly isotropic in the sky and no obvious
counterpart has been seen in the arrival directions of the highest
energy events, in particular the world record Fly's Eye
event~\cite{ES95} or the highest energy Auger
event~\cite{KL08}. Departures from isotropy have been reported
recently by the Pierre Auger Observatory through an apparent
correlation of the arrival directions of cosmic rays with $E>6\times
10^{19}\,$eV with nearby extragalactic matter within $70\,$Mpc. The
first data set rejected isotropy at the $3\sigma$ level~\cite{Aea08b}.
It has been shown to agree reasonably well with the distribution of
large scale structure in the nearby
Universe~\cite{Gea08,KW08,KL08,KT09}. Integrating more recent data,
the evidence has slightly weakened although the signal has remained
roughly at the $3\sigma$ level. Likelihood tests of this distribution
suggests a correlation with nearby large scale structure if one also
allows for isotropy in a substantial fraction of arrival
directions~\cite{Aea10PAOb}. Such evidence for anisotropy has however
not been recorded by the HiRes experiment in the Northern sky, which
rejects correlation with large scale structure at a 95\% confidence
level~\cite{Aea10b}.  Whether or not the discrepancy between these
experiments is real remains to be proven: for one, the significance
level of either claim is not so high; secondly, the energy scales of
both experiment differ by some $20\,$\%, which translates in a
significant difference in terms of horizon distance (hence anisotropy
signal), because the horizon distance evolves rapidly with energy
close to the GZK cut-off.

\begin{figure}
\includegraphics[width=23pc]{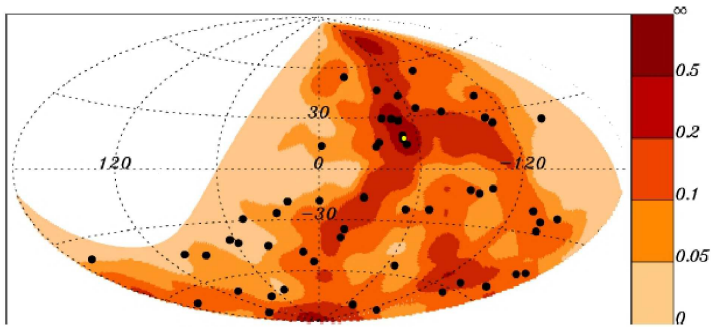}
\begin{minipage}[b]{14pc}
  \caption{Expected flux above $5.5\times 10^{19}\,$eV for the Pierre
    Auger Observatory, assuming that the sources are distributed as
    the galaxies of the 2MRS catalog. Black dots: Auger events; yellow
    dot: location of Cen~A. Adapted
    from~\cite{Aea10PAOb}.\label{fig:skymap} }
\end{minipage}
\end{figure}

Most of the anisotropy pattern seen by Auger results from an apparent
clustering of arrival directions around Centaurus~A, see
Fig.~\ref{fig:skymap}. Being the nearest giant radio-galaxy and a
potential site of ultrahigh energy cosmic rays, Cen~A has soon enough
been charged of accelerating particles up to $10^{20}\,$eV. This
however is too bold a step, which is not supported by theoretical and
phenomenological arguments, as discussed above and further below, see
also ~\cite{LW09}.  Cen~A happens to lie in the direction to one the
strongest concentrations of matter in the nearby Universe: the
Centaurus supercluster at $\sim 50\,$Mpc and the Shapley supercluster
at $\sim 200\,$Mpc; furthermore Cen~A does not lie far from the zenith
of the Pierre Auger Observatory. Therefore, if sources are distributed
as the large scale structure, and the magnetic deflection rather weak,
a cluster of events in that direction of the sky comes at no real
surprise~\cite{KW08}.

\subsection{Cosmic ray deflection}\label{sec:crdef}
A cosmic ray suffers deflection in the intervening extragalactic and
Galactic magnetic fields. For detailed studies on the amount of
deflection for protons and nuclei in the Galactic magnetic field, see
e.g.~\cite{AMS06,TS08,GKSS10,TIY12}. The amount of deflection varies
from direction to direction, in a way which depends sensitively on the
poorly known Galactic field, but roughly, the order of magnitude is
$1^\circ$ for protons at $10^{20}\,$eV, less at high Galactic
latitudes and obviously $Z$ times more for nuclei of charge $Z$. At
lower energies, nuclei arrival directions can suffer from lensing in
the Galactic magnetic field~\cite{HMR02}.

The strength and distribution of large scale extragalactic magnetic
fields are very poorly known, mostly because the origin of these
magnetic fields is unknown, and because their evolution during the
formation of large scale structure is far from trivial,
notwithstanding all the possible sources of magnetic pollution in the
late Universe. The impact of such magnetic fields on the transport of
ultrahigh energy cosmic rays can be described roughly as follows.

At the highest energies, the particles only experience the rare
localized regions of sufficiently intense magnetic field: the gyration
radius $r_{\rm L}\,\simeq\, 10\,{\rm Mpc}\,Z^{-1}E_{20}B_{-8}^{-1}$ is
indeed much larger than the coherence length $\lambda_B$ of
intergalactic turbulence, which is bounded to
$\lesssim100\,$kpc~\cite{WB99}, hence the deflection $\delta\theta
\sim l\lambda_B/r_{\rm L}^2\,\ll\,l/r_{\rm L}$ in a structure of size
$l$. Consequently, the transport of such cosmic rays can be described
by a series of stochastic interactions with the peaks of the
intergalactic magnetic field distribution, see~\cite{KL08} for a
discussion. Such regions typically involve the filaments of large
scale structure, which may be contaminated up to $B_{-8}\sim1$,
corresponding to $\sim10\,$\% of equipartition. The direction of the
cosmic ray thus follows a random walk and arrives on the detector with
a typical spread of order of a few degrees for protons of
$10^{20}\,$eV on a distance scale of $100\,$Mpc. The dependence on the
parameters characterizing the magnetic field distribution are given in
Ref.~\cite{KL08}. The expected angular deflection is of course 26
times larger for iron nuclei of the same energy. The corresponding
time delay between the arrival of a photon emitted at the same time as
the cosmic ray and that of the cosmic ray is of the order of
$10^4\,$yrs, which can explain why transient sources are not seen in
the arrival direction of these highest energy particles~\cite{WME96}.

At lower energies, the accumulated angular deflection increases, and
it actually blows up rapidly below the GZK threshold, as a consequence
of the rapidly increasing horizon distance, which sets the typical
distance to the source, i.e. the distance from which most of the flux
is collected. This geometrical effect, combined with the decreasing
gyration radius, implies very large deflections, so that one should
expect a roughly isotropic sky below the GZK threshold, as observed
indeed.

At such energies and on such path lengths, the cosmic rays start to
diffuse in the extra-galactic magnetic field. As the energy of the
cosmic ray decreases further, it becomes sensitive to regions of
weaker and weaker magnetic field, which have a higher filling factor
in the large structure. An interesting consequence is the following:
if the diffusion time scale from the closest cosmic ray source exceeds
a Hubble time, the flux is suppressed. This leads to a low-energy
cut-off in the extragalactic cosmic ray flux, expected around
$10^{17}-10^{18}\,$eV for protons, i.e. in an interesting region to
explain the transition between Galactic and extragalactic cosmic
rays~\cite{L05,AB05}.

\subsection{Anisotropy vs chemical composition}\label{sec:chem-aniso}
Given the numerous unknowns on the configuration of the Galactic and
extragalactic magnetic fields, as well as on the charge of the
particles, it is an almost impossible exercise to backtrack the
position of the sources from the arrival directions in the sky with
existing data. However, one may test the chemical composition on the
sky by quantifying the degree of anisotropy in a particular direction
as a function of energy, using the fact that sources producing heavy
nuclei of charge $Z$ at an energy $E$ must produce a similar
anisotropy pattern at energies $E/Z$ through the proton component that
is accelerated along with heavier nuclei. This test is discussed in
detail in Ref.~\cite{LW09}. The anisotropy pattern is generally
measured as an excess number of events in a region of the sky
relatively to an isotropic background.  Although the isotropic
background "noise" increases when one goes from $E$ to $E/Z$, the
signal-to-noise ratio of the anisotropy pattern actually increases
significantly, because the number of events in the anisotropy region
also increases. This prediction does not depend on the modelling of
astrophysical magnetic fields as it only relies on the property that
protons of energy $E/Z$ follow the same path in the intervening
magnetic fields and produce the same angular image as heavy nuclei of
charge $Z$ and energy $E$, with whom they share the same magnetic
rigidity.

It is useful to take as an example the observed clustering around
Centaurus~A, $12$ events located within $18^\circ$ out of a total of
$58$ events above $55\,$EeV for $2.7$ expected for isotropic arrival
directions, as of 2009. Then, assuming that the sources inject protons
and iron nuclei with a chemical composition similar to that of
Galactic cosmic rays (at their source), a standard $-2$ powerlaw
spectrum and maximal energy $3\times Z\,$EeV, so that only iron nuclei
are observed at GZK energies, one would expect an anisotropy signal
many times larger (in terms of signal-to-noise ratio) at $2-3$EeV than
is observed at GZK energies, see ~\cite{LW09}. One may indeed relate
the signal-to-noise ratios of the anisotropy pattern, i.e.
$\Sigma_p(>E/Z)$ for protons at $E/Z$ and $\Sigma_Z(>E)$ for heavy
nuclei at energy $E$ through
\begin{equation}
\Sigma_p(>E/Z)=\alpha_{\rm loss}\,\frac{N_p}{N_Z}\,Z^{(1-s_{\rm
    obs})/2}\,\Sigma_Z(>E)\ ,
\end{equation}
where $N_p/N_Z$ represents the abundance ratio of proton to nuclei at
the source, $\alpha_{\rm loss}>1$ characterizes the influence of
photodisintegration of heavy nuclei, and $s_{\rm obs}\simeq 2.7$ is
the spectral index of the observed all particle spectrum. The above
assumes for the injection spectrum a generic powerlaw function of
magnetic rigidity. For a solar abundance pattern, $Z^{(1-s_{\rm
    obs})/2}N_p/N_Z$ takes values of order $(100, 600, 2000)$ for
$Z=8$ (CNO group), $Z=14$ (Si group) and $Z=26$ (Fe group).

This test has been applied to real data in Ref.~\cite{Aea11} and no
significant anisotropy has been observed at low energies. If the
anisotropy observed at GZK energies is not a statistical accident,
this result appears difficult to reconcile with a heavy composition at
the highest energies. It would indeed require an extraordinarily high
metallicity.  On top of this, the measured all sky cosmic ray
composition indicates a significant fraction of protons at low
energies~\cite{Aea10,Aea10PAO}. Of course, if the clustered events at
high energies are protons, the absence of anisotropies at lower
energies follows naturally from magnetic smearing.

The radio-galaxy Cen~A does not possess the required characteristics
to accelerate protons to $10^{20}\,$eV, as discussed in
Section~\ref{sec:cand}: the jet kinetic power in Cen~A $L_{\rm
  jet}\,\simeq\,2\times 10^{43}\,$erg/s and the fit of the spectral
energy distribution of the nucleus indicates
$L_B\,\sim\,10^{42}\,$ergs/sec~\cite{Cea01}. Based on this and other
arguments~\cite{CLP02,LW09}, one finds that the maximal energy in
Cen~A does not exceed $\sim 3\times Z\times 10^{18}\,$eV. One might
argue that Cen~A can accelerate heavy nuclei to GZK energies, but this
would be in strong conflict with the absence of anisotropies at EeV
energies. Cen~A thus does not appear to be a significant source of
ultrahigh energy cosmic rays.  It thus seems much more conservative to
interpret the clustering around Cen~A as a signature of the
distribution of sources in the large scale structure.  Although, if
gamma-ray bursts (and/or magnetars) accelerate protons to ultrahigh
energies, then the last gamma-ray bursts in the host galaxy of Cen~A
would contribute to the flux from Cen~A, due to the rescattering of
the emitted particles on the giant lobes of this
radio-galaxy~\cite{LW09}, with a contribution that could acount for a
few events in the current Pierre Auger data set.

\subsection{Multi-messengers}
Finally, one hope to derive indirect information of the source is to
collect high energy photons or neutrinos associated to the
acceleration and interaction of primary ultrahigh energy cosmic rays,
either in the source or en route to the detector. One most salient
result in this area is the non observation of PeV neutrinos from
gamma-ray bursts by the Ice Cube experiment~\cite{IC12}. The current
upper limit lies very close to the model prediction of Waxman \&
Bahcall~\cite{WB97,HBW12,Hea12}. If no neutrino were observed, it
would not however exclude gamma-ray bursts as sources of ultrahigh
energy cosmic rays, because acceleration might take place elsewhere in
the source, with different secondary neutrino signatures.

An interesting result is the possible explanation of the TeV
``excess'' of remote blazars as the secondary photons of very high
energy cosmic rays accelerated in these blazars~\cite{E10} (and
references therein). It is intriguing however, because this model
requires a very high injected energy in cosmic rays~\cite{RDF12}, all
the more so when realistic deflection in extragalactic magnetic fields
is accounted for~\cite{MDTM12}.

The expected flux of very high energy photons from ultrahigh energy
cosmic ray sources is generally quite low, see e.g. ~\cite{CW93,M12}
for transient sources, ~\cite{A02,Fea04,GA05,Kea11,PKA11} for
continuous sources. It may result from two effects: one is the
production of a very high energy photon that cascades down to the
GeV-TeV range through successive pair conversion and inverse Compton
in the diffuse radiation fields. The observation of such photons
requires very low extragalactic magnetic fields, as otherwise the
electron/positron would be deflected out of the beam before
transferring its energy to a background photon. Typically, this
implies $B\lesssim 10^{-13}\,$G; because the cascade development is
quite fast, such cascades could take place in the weakly magnetized
voids of large scale structure~\cite{CW93,Kea11}. The other process is
the production of an electron that radiates in synchrotron in a region
of sufficient field strength, before transferring its energy to a
background photon~\cite{GA05}. This channel is particularly
interesting because it may lead to a non ambiguous signature, under
the form of a degree size halo around the source in the $10-100\,$GeV
range, with mild dependence on the photon energy in that range. Both
channels are expected to contribute at a roughly similar level. To
become detectable, a continuous source should output some
$10^{46}\,$erg/s in cosmic rays at $1\,$Gpc, although the budget
requirement becomes less at high redshift due to the evolution of
diffuse backgrounds~\cite{PKA11}.

Finally, diffuse backgrounds are to be expected, among which the
diffuse neutrino background at $\sim100\,$PeV, associated to GZK
interactions of cosmic ray protons. Obviously, if the composition
becomes heavy close to the GZK threshold, the expected flux
diminishes, down to quite pessimistic estimates~\cite{KAO10}.

\section{Conclusions and perspectives}\label{sec:concl}
To summarize, the development of large scale cosmic ray detectors has
led to substantially increased statistics at the highest energies and
most particularly, to the discovery of the cut-off at $6\times
10^{19}\,$eV, the expected location for the GZK cut-off beyond which
the Universe becomes opaque to cosmic ray nuclei. The recent data of
the Pierre Auger Observatory has also brought evidence for an
increasingly heavier composition above the ankle, in apparent conflict
with the chemical composition reconstructed by the HiRes and now the
Telescope Array experiment. Resolving this issue becomes a priority as
different chemical compositions imply different interpretations of
existing data.  For instance, in Sec.~\ref{sec:cand} it was shown that
the magnetic luminosity of the source of ultrahigh energy cosmic rays
must verify $L_B\gtrsim 10^{45}Z^{-2}\,$erg/s in order to accelerate
particles of charge $Z$ up to $10^{20}\,$eV. Whether one is dealing
with protons or iron nuclei then leads to different conclusions. With
respect to protons, only a few types of astrophysical objects seem
capable of achieving acceleration to $10^{20}\,$eV, while the bound is
substantially weaker for iron nuclei. Likewise, the sky map of arrival
directions takes different appearances depending on the nature of the
primary: for $10^{20}\,$eV protons, the expected deflection is on the
order of a few degrees, while for iron nuclei, it is $\sim26$ times
larger. In the former case, one should thus expect to see a
correlation with the large scale structure at the highest energies. In
the latter case of heavy nuclei primaries, the exact pattern of
anisotropies, if any, is very difficult to predict with accuracy given
the existing uncertainties on the configuration of the galactic
magnetic field.

To constrain the chemical composition further, one may use the
anisotropy pattern in the sky as a function of energy. This method has
been discussed in Sec.~\ref{sec:chem-aniso} and applied to the
apparent clustering reported by the Pierre Auger Observatory around
the Cen~A source. The absence of anisotropies at EeV energies then
suggests that the clustered events at GZK energies are mostly protons;
otherwise one would have expected to observe a much stronger
anisotropy signal at EeV energies, that should be associated with the
protons of a same magnetic rigidity as the hypothesized iron nuclei at
GZK energies.

Regarding the perspectives in this field, one can only consider
extreme cases to extrapolate the current situation, given the apparent
conflict on the measurement of the chemical composition. Assume first
that ultrahigh energy cosmic rays are mostly protons. Then one needs
to understand of course why the Pierre Auger Observatory reports a
heavier composition beyond the ankle. Only a few candidates seem
capable of accelerating protons to ultrahigh energies, and the
apparent lack of counterpart in the arrival directions of the highest
energy events, together with the expected small angular deflection
then suggests that the source is a transient object, such as gamma-ray
burst or a magnetar, camouflaged in the large scale structure. To make
progress towards source identification, one needs to acquire more
statistics at the highest energies to obtain a more accurate spectrum,
possibly isolate clusters of events in the sky and more generally,
search for the expected correlation with the large scale
structure. This will be the goal of planned and future experiments
such as JEM-EUSO~\cite{E11} or an extension of the Pierre Auger
Observatory~\cite{Aea09}. In parallel, one should search for
signatures of ultrahigh energy cosmic ray acceleration in
multi-messenger signals of gamma-ray bursts and/or magnetars. As a
noteworthy example, the Ice Cube experiment is now probing the PeV
neutrino signal at the Waxman - Bahcall bound predicted in some
scenarios of ultrahigh energy cosmic ray acceleration in gamma-ray
bursts~\cite{IC12}.

If the ultrahigh energy cosmic rays are mostly heavy nuclei, such as
iron, the situation becomes unfortunately much more complex in terms
of phenomenology: anisotropies should be absent or at the best weak,
with a pattern that is hard to predict; theory cannot help much
because the pool of source candidates is substantially larger, due to
the greater facility to accelerate nuclei to $10^{20}\,$eV; finally,
the possibility of detecting multi-messenger signals is considerably
reduced due to the smaller Lorentz factor comparatively to protons of
a same energy.

In between these two extremes lies a continuum of scenarios with mixed
composition models, multi-source models... Interestingly, if protons
are present up to ultrahigh energies, one recovers a situation in
which only a few objects can subscribe as potential sources and in
which anisotropies might be detectable, although any heavy nuclei
component inputs background noise on such anisotropies. One should
then focus the search on the highest rigidity particles in Nature,
rather than the highest energy ones, as they must come from the most
extreme sources and they likely lead to the most optimistic prospects
in terms of source reconstruction.

Which scenario best describes the data is currently (and will remain
for some time) the discussion of lively debates.\bigskip

\noindent{\bf Acknowledgments:} This work has been supported in part
by the PEPS-PTI Program of the INP (CNRS).


\end{document}